\documentclass[a4paper,11pt,dvips]{article}
\usepackage{graphicx}
\usepackage{amsmath}
\usepackage{amssymb}
\usepackage{latexsym}
\usepackage[colorlinks]{hyperref}
\usepackage{color}
\usepackage{float}
\usepackage{cite}
\usepackage{makeidx}
\usepackage{colortbl}

\newcommand{\bra}{\begin{array}}
\newcommand{\era}{\end{array}}
\newcommand{\beq}{\begin{equation}}
\newcommand{\eeq}{\end{equation}}
\newcommand{\bqr}{\begin{eqnarray}}
\newcommand{\eqr}{\end{eqnarray}}

\def\BC{\bb C}
\def\_\BC{\bbi C}

\def\no2 {{\textstyle{n\over 2}}}

\newcommand{\lb}{\label}

\begin{document}
\begin{titlepage}
\setcounter{page}{1}
\renewcommand{\thefootnote}{\fnsymbol{footnote}}

\begin{flushright}
\end{flushright}

\vspace{5mm}
\begin{center}

{\Large \bf {Multibands Tunneling in AAA-Stacked Trilayer
Graphene}}

\vspace{5mm}
{\bf Ilham Redouani}$^{a}$,
 {\bf Ahmed Jellal\footnote{\sf ajellal@ictp.it --
a.jellal@ucd.ac.ma}}$^{a,b}$, {\bf Abdelhadi Bahaoui}$^{a}$ and {\bf Hocine Bahlouli}$^{b,c}$

\vspace{5mm}

{$^{a}$\em Theoretical Physics Group,  
Faculty of Sciences, Choua\"ib Doukkali University},\\
{\em PO Box 20, 24000 El Jadida, Morocco}

{$^b$\em Saudi Center for Theoretical Physics, Dhahran, Saudi Arabia}

{$^c$\em Physics Department,  King Fahd University
of Petroleum $\&$ Minerals,\\
Dhahran 31261, Saudi Arabia}


\vspace{3cm}

\begin{abstract}

We study the electronic transport through {\it np} and {\it npn} junctions
for AAA-stacked trilayer graphene. Two kinds of
gates are considered where the first is a single gate and the second is a double gate.
After obtaining the solutions for the energy spectrum, we use the
transfer matrix method to determine the three transmission
probabilities for each individual cone $\tau=0, \pm 1$.
We show that the quasiparticles in AAA-stacked trilayer graphene are not only
chiral but also labeled by an additional cone index $\tau$.
The obtained bands are composed of three Dirac cones
that depend on the chirality indexes. We show that
there is perfect transmission for normal or near
normal incidence, which is a manifestation of the Klein tunneling effect.
We analyze also 
the corresponding
total conductance, which is defined as the sum of the conductance
channels in each individual cone. 
Our results are numerically discussed and compared with those obtained
for ABA- and ABC-stacked trilayer graphene.

\end{abstract}
\end{center}

\vspace{3cm}


\end{titlepage}


\section{Introduction}

During the last decade, the physics of single layer graphene and
stacks of graphene layers has emerged as a fertile research area
\cite{Novoselov, Novoselov2, Zhang}. This is due to its unusual
electronic properties, that may be useful in the design of new
electronic devices \cite{Berger,Bunch,Novoselov3}. 
Among them, we cite the   linear
dispersion relation of a single graphene layer, manifestation of the Klein tunneling,
\cite{Novoselov2,Zhang}, chiral parabolic bands in bilayers
\cite{Novoselov4} and the possibility of confining charge to the
surface in systems with a multilayer graphene \cite{Morozov}.
Furthermore, it has been observed in different works \cite{Partoens, Koshino,
Nilsson, Avetisyan, Koshino2, DasSarma, Jung, DeAndres, Munoz} that the properties of multilayer graphene materials
depend on their stacking order and the number of layers.
Most of these works have been devoted to the few-layer
graphene materials with Bernal (ABA) and rhombohedral (ABC)
stacking order \cite{Duppen,Duppen2}. For the rhombohedral form,
the Klein tunneling depends on the staking order \cite{Bala,
Duppen3} and
is absent for the Bernal
stacking.


Recently a new stable multilayer graphene with AAA-stacking order
has been experimentally realized \cite{Quhe12}.
In such system, each sublattice in the top layer is located directly
above the same one in the bottom layer. Due to this staking order,
the AA-stacked bilayer graphene (BLG) has a special low energy band
structure.
It is
just the double copies of single layer
graphene bands shifted up/down by the interlayer coupling
$\gamma= 0.2\ eV$ \cite{Tabert} and  
is also different from that corresponding to the AB-stacked BLG. Due to this special band structure, the AA-stacked BLG
shows many interesting properties which are different from that of
single layer and also not been observed in the other
graphene-based materials \cite{Tabert,Ando2, Hsu,Prada,Brey,
Mohammadi2}.

There has been a growing interest in the study of the
tunneling problem of charge carriers in trilayer graphene systems
including ABA- and ABC-stacked trilayer graphene (TLG)
\cite{Duppen,Duppen2}. In the present work, 
we consider the AAA-stacked TLG
as
schematically shown in Figure \ref{Fig.barrier}(a),
which is composed of three
single layers, each sublattice in a top layer is located directly
above the same one in the bottom layer.
The unit cell of an AAA-stacked consists of 6 inequivalent carbon
atoms with two atoms for each layer.
We study
the electronic transport through {\it np} and {\it npn} junctions for AAA-stacked TLG
where
two kinds of
gates will be considered such that the first causes an equal potential shift $V$ for all three
layers and the second induces an interlayer potential difference
$\delta$ between neighboring layers. We show that the quasiparticles in
AAA-stacked TLG are not only chiral but also labeled
by an additional cone index $\tau$. We obtain 
band structures that are composed of three Dirac cones and depending on
the cone indexes and the chirality indexes. Our theoretical model
is based on the well established tight binding Hamiltonian
\cite{Mohammadi}.
{The single and double gates act
as a boundary for which we calculate the three transmission
probabilities as  function of the angle of incidence and
Fermi energy of the incident electron for different configurations
of the devices.} Subsequently, 
we numerically evaluate 
the total conductance
and underline the basic features of our system. 


The rest of the paper is organized as follows. In section 2, we
formulate our model by setting the Hamiltonian system used to
describe AAA-stacked TLG. We explore the
mirror reflection symmetry of the lattice in the plane
of its central layer to determine the solutions of the energy spectrum
in each layer ($\tau=-1, 0,1$). Later on, we present the
formalism, indicate the different propagating modes, define the
four different cases for transmission for the six-band model and
explain only the three possible transmission probabilities. In
section 3,  using the transfer matrix at boundaries together with
the incident, transmitted and reflected currents we end up with
three transmission probabilities. Next, we numerically discuss the
obtained transmissions for each individual cones ($\tau=0, \pm 1$)
for {\it np} and {\it npn} junctions to underline the behavior of
our system. {In section 4, we show
the numerical results for the conductance and investigate the
contribution of each transmission channel}. Finally, we conclude
our work
and emphasize our main results. 

\section{Tight binding formalism}

\begin{figure}[h!]
 \centering
 \includegraphics[width=4cm, height=3cm]{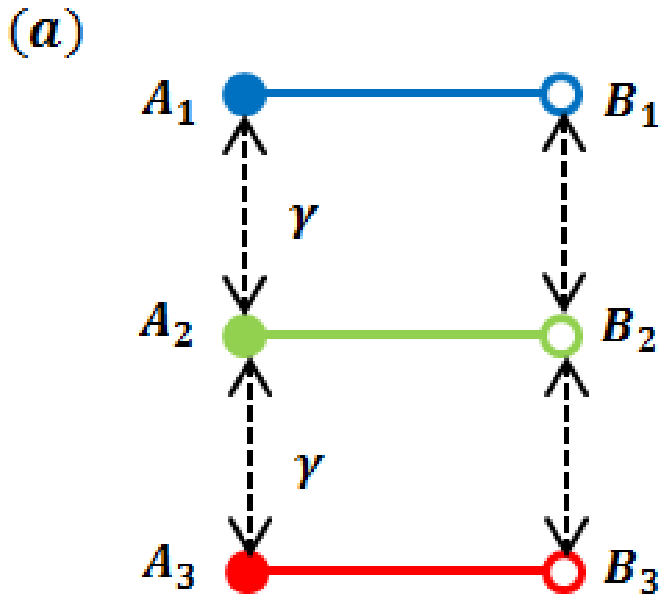}
 \ \ \
 \includegraphics[width=7cm, height=3cm]{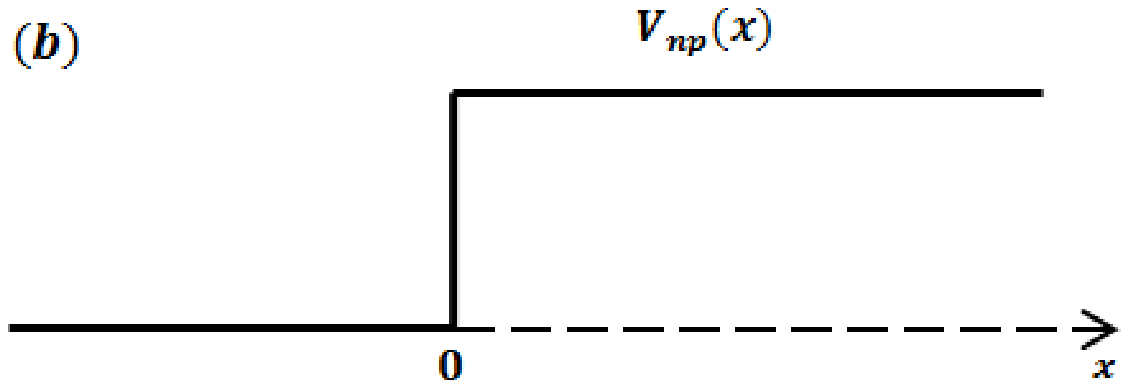}
\ \ \
 \includegraphics[width=8.5cm, height=3cm]{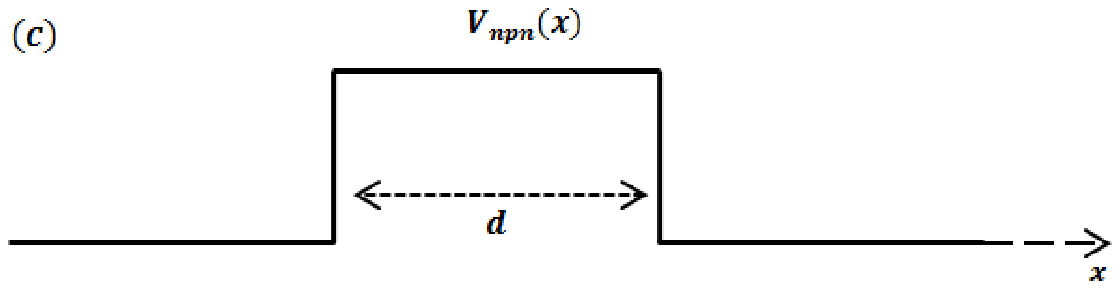}
 \caption{\sf (a): Schematic representation of the lattice structure of the AAA-stacked trilayer graphene with (A,B) atoms within the
same layer and $\gamma$ is the interlayer hopping energy. (b):
Diagram for the {\it np} junction of height
$V_{{np}}(x)$. (c): Diagram for the {\it npn} junction of height
$V_{{npn}}(x)$ and width $d$.}\label{Fig.barrier}
\end{figure}

We consider a system consisting of three layers of graphene having
the AAA-stacking structure. The unit cell consists of six
atoms labeled $A_1$, $B_1$ in the top layer, $A_2$, $B_2$ in the
central layer and $A_3$, $B_3$ in the bottom layer as depicted in
Figure \ref{Fig.barrier}(a). Each carbon atom of the bottom
(center) layer is located above the corresponding atom of the
center (top) layer, respectively and they are bound by an
interlayer coupling energy $\gamma= 0.2\ eV$. For the considered
system, we define two different potential profiles where the first causes
a potential shift $V$ equal for all three layers and the second one
induces an interlayer potential difference $\delta$ between
neighboring layers.

\begin{figure}[h!]
 \centering
 \includegraphics[width=6cm, height=5cm]{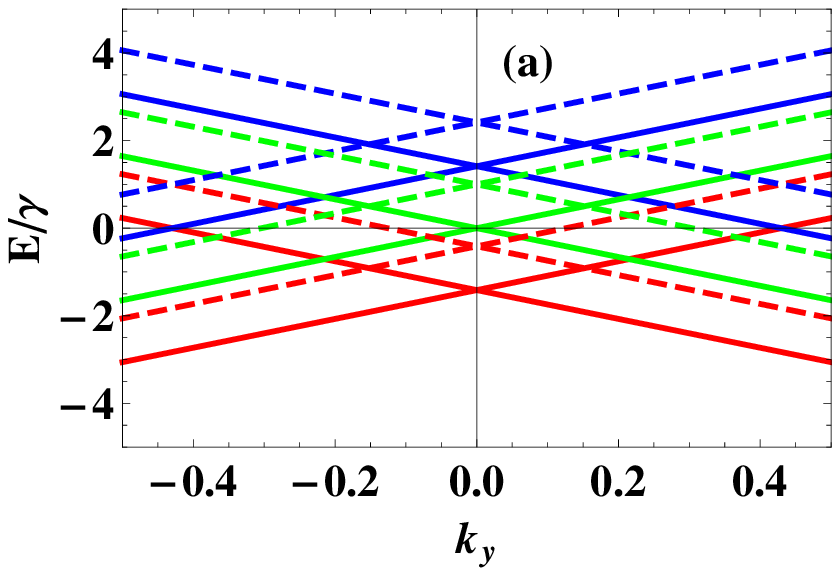}
\ \ \
 \includegraphics[width=6cm, height=5cm]{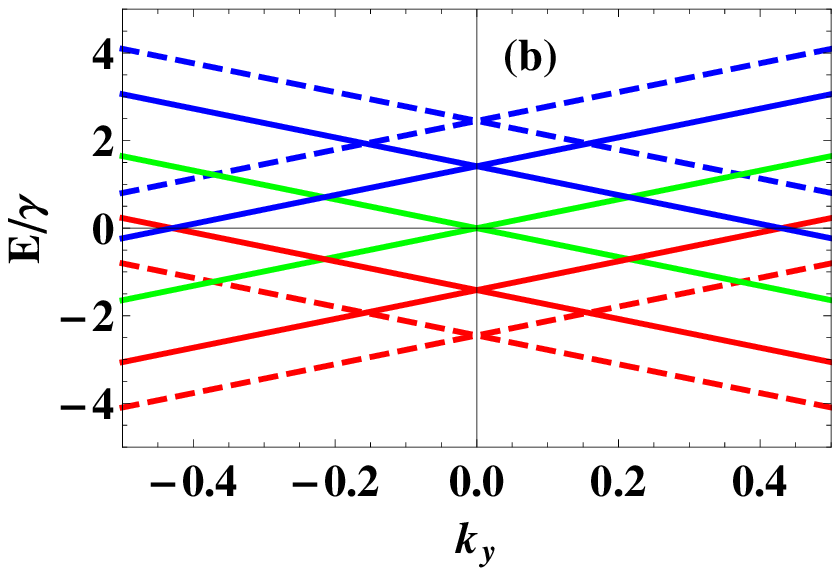}
 \caption{\sf The six energy eigenvalues inside and outside the barrier regions as
 function of the momentum $k_y$. The red, green and blue lines correspond to cone indexes
 $\tau=-1,0,1$, respectively, with $V=\delta=0$. (a):  $V=1 \gamma$ and
$\delta=0$ (dashed lines). (b): $V=0$ and $\delta=2 \gamma$
(dashed lines).}\label{Fig.energy}
\end{figure}

The Dirac fermions are scattered by an {\it np} junction and an {\it npn} junction
along the \textit{x}-direction. 
Therefore, the charge carriers in AAA-stacked trilayer is described by the following six-band Hamiltonian
\begin{equation}\label{hamil1}
{H_{(AAA)}=\left( \begin{array} {ccc} (V+\delta)\mathbb{I}_2 +
 v_F \overrightarrow{\sigma}\cdot\overrightarrow{p}
 & \gamma\mathbb{I}_2  & 0\\
\gamma\mathbb{I}_2 & V \mathbb{I}_2 +
 v_F \overrightarrow{\sigma}\cdot\overrightarrow{p} & \gamma\mathbb{I}_2 \\
 0 & \gamma\mathbb{I}_2 & (V-\delta) \mathbb{I}_2 +
 v_F \overrightarrow{\sigma}\cdot\overrightarrow{p}
\end{array}\right)}. 
\end{equation}
The eigenstates of $H_{(AAA)}$ 
are the six component spinors $\Psi(x,y)=\left(\Psi_1,\ \Psi_2,\
\Psi_3 \right)^T$, where
$\Psi_{1,2,3}=\left(\Psi_{A_{1,2,3}},\Psi_{B_{1,2,3}} \right)^T$
are the envelope functions associated with the probability
amplitudes of the wave functions on the $A_{1,2,3}$ and
$B_{1,2,3}$ sublattices of the three layers. In
\eqref{hamil1}, $\overrightarrow{\sigma} =(\sigma_x, \sigma_y)$ is
a vector of Pauli matrices, $v_F$ is the Fermi velocity in
monolayer graphene, $\overrightarrow{p}=(p_x,p_y)$ is the momentum, $V$ is a general potential term, $\delta$ corresponds to
an externally induced interlayer potential difference and $\gamma$
describes the interlayer hopping parameter.
Exploiting
mirror reflection symmetric of the lattice in the plane of its
central layer, we perform a unitary transformation to a basis
of the spinor components by combining the atomic wave
functions $\Psi_{A_1}$ with $\Psi_{A_3}$ and $\Psi_{B_1}$ with
$\Psi_{B_3}$.
This operation
 transforms $\Psi(x,y)$ to
$\Psi(x,y)=\left(\Psi^+,\Psi_{2},\Psi^-\right)^T $, where
$\Psi^\pm=\left(\Psi_{A}^\pm,\Psi_{B}^\pm\right)^T$,
$\Psi_{A}^\pm=\left(\Psi_{A_1}\pm\Psi_{A_3}\right)/\sqrt{2}$ and
$\Psi_{B}^\pm=\left(\Psi_{B_1}\pm\Psi_{B_3}\right)/\sqrt{2}$.
With this, the Hamiltonian \eqref{hamil1} is transformed into
\begin{equation}\label{TrH}
H_{(AAA)}^{'}=\left( \begin{array} {ccc} V \mathbb{I}_2 + v_F
\overrightarrow{\sigma}\cdot\overrightarrow{p}
 & \sqrt{2}\gamma \mathbb{I}_2 & \delta \mathbb{I}_2\\
\sqrt{2}\gamma \mathbb{I}_2 & V \mathbb{I}_2 +
v_F \overrightarrow{\sigma}\cdot\overrightarrow{p} &  0 \\
\delta \mathbb{I}_2 & 0 & V \mathbb{I}_2 + v_F
\overrightarrow{\sigma}\cdot\overrightarrow{p}
\end{array}\right).
\end{equation}
The above form of the Hamiltonian consists of a $2\times 2$
monolayer-like (bottom, right) and a $4\times 4$ bilayer-like
(top, left) blocks that are connected by the parts responsible for
the interlayer potential difference $\delta$. The two blocks
result in a superimposed linear from the monolayer and bilayer
spectrum near the Dirac point as shown  in \ref{Fig.energy}. In
that case, electrons propagating in AAA-stacked TLG can propagate through
two different modes, one monolayer-like and one AA-stacked
bilayer-like mode. As long as the mirror symmetry remains intact,
both modes will not interact and scattering between them is
prohibited. Here we shall study the problem of the tunneling of
electron through a {\it np} junction (Figure \ref{Fig.barrier}(b))
and {\it npn} junction
(Figure \ref{Fig.barrier}(c)) 
\begin{equation} \label{eq 2}
V_{{np}}(x)=\left\{\begin{array}{lll}
{V \mathbb{I}_2+\Delta} & \mbox{if} & {x > 0 } \\
{0} & \mbox{if} & {x < 0} \end{array}\right.,
\qquad
V_{{npn}}(x)=\left\{\begin{array}{lll}
{V \mathbb{I}_2+\Delta} & \mbox{if} & {0 < x < d } \\
{0} & \mbox{if} & {x <0, \ x > d}
\end{array}\right.
\end{equation}
where $V$ is the height of the potential, which corresponds to the single
gate term, and
the parameter $\Delta=\text{diag}\{\delta,\delta,0,0,-\delta,-\delta\}$
describes the effect of a double gate.  The potential is $V$ for
the center layer and $V \pm \delta$ for the top and bottom layers.

To proceed further, let us set the length scale
$\eta=\frac{\hbar v_F}{\gamma}\thickapprox 3.29~nm$ 
as well as employ the dimensionless quantities
to simplify the notation such that the energy terms can be parameterized
by interlayer coupling $\gamma$, then $E \longrightarrow
\frac{E}{\gamma}$, $V \longrightarrow \frac{V}{\gamma}$,
$\Delta\longrightarrow \frac{\Delta}{\gamma}$, $k_y \longrightarrow \eta k_y$
and $x \longrightarrow \frac{x}{\eta}$. The system is infinite
along the \textit{y}-direction so that $[H_{(AAA)}^{'},p_y]=0$ and hence
we can write $\Psi(x,y)=e^{ik_yy}\psi(x)$. The Hamiltonian
\eqref{TrH} used together with the spinor $\Psi(x)$ in the eigenvalue equation
$H_{(AAA)}^{'}\Psi(x)=E\Psi(x)$ to get  six linear
differential equations of the from
\begin{subequations}\label{eq3a}
\begin{eqnarray}{}
&&{ -i (\partial_x+k_y)\psi_{B}^{+}+\sqrt{2}\psi_{A_2}}={(E-V)\psi_{A}^{+}-\delta\psi_{A}^{-}}\\
&&{-i (\partial_x-k_y)\psi_{A}^{+}+\sqrt{2}\psi_{B_2}}={(E-V)\psi_{B}^{+}-\delta\psi_{B}^{-}}\\
&&{-i (\partial_x+k_y)\psi_{B_2}+\sqrt{2}\psi_{A}^{+}}={(E-V)\psi_{A_2}}\\
&&{-i (\partial_x-k_y)\psi_{A_2}+\sqrt{2}\psi_{B}^{+}}={(E-V)\psi_{B_2}}\\
&&{-i (\partial_x+k_y)\psi_{B}^{-}}={(E-V)\psi_{A}^{-}-\delta\psi_{A}^{+}}\\
&&{-i\ (\partial_x-k_y)\psi_{A}^{-}}={(E-V)\psi_{B}^{-}-\delta\psi_{B}^{+}}.
\end{eqnarray}
\end{subequations}
Combining the above equations by eliminating the unknowns one
at time to end up with the second order differential equation
\begin{equation}
\left[\partial_{x}^2+\left(k_{x}^\tau\right)^2\right]\psi_{B}^{+}(x)=0
\end{equation}
where the wave vectors along the \textit{x}-direction are defined
by
\begin{equation}
{k_{x}^\tau}={\sqrt{-k_{y}^2 + \left(E-V-\tau\sqrt{\delta^2+2}\right)^2}}.
\end{equation}
This form of the wave vectors consists of monolayer-like
($k_{x}^0$) and AA-stacked bilayer-like ($k_{x}^\pm$), where
$\tau$ is the cone index such that $\tau=0$ for the center cone and
$\tau=-1$ ($\tau=+1$) for the bottom (top) cone. The energy bands
corresponding to the wave vectors $k_{x}^{\tau}$ read as
\begin{equation}\label{energybands}
E^{s,\tau}= V+\tau\sqrt{\delta^2+2}+s
\sqrt{(k_{x}^{\tau})^2+(k_{y})^2}.
\end{equation}
The energy eigenvalues outside the barrier are then given by
\begin{equation}\label{energybands0}
E^{s_0,\tau}= \tau \sqrt{2}+s_0
\sqrt{(k_{x_0}^{\tau})^2+(k_{y})^2}.
\end{equation}
Here $s$ and  $s_0$ are the chirality indexes of a quasiparticle in the barrier
and outside the barrier regions, respectively. Here $s =+1$ (or $s_0 =+1$) and
$s =-1$ (or $s_0 =-1$) for electron-like and hole-like particles,
respectively. The wave vector $k_{x_0}$ corresponding to this energy
band $E^{s_0,\tau}$ is given by
${k_{x_0}^\tau}={\sqrt{-k_{y}^2+\left(\eta\right)^{-2}\left(E-\tau
\sqrt{2}\right)^2}}$.

Plot of the energy bands inside and outside the barrier regions, as
defined in \eqref{energybands} and \eqref{energybands0},
as a function of the momentum $k_y$ are shown in Figure
\ref{Fig.energy}. In this Figure, $\tau=-1$ (red line) correspond
to the bottom cone, $\tau=0$ (green line) for the center cone and
$\tau=1$ (blue line) for the top cone.
Whereas, the energy bands, for $\delta=0$, of the AAA-stacked
trilayer graphene are linear and just like three copies of the
monolayer band structure, where the two bands ($\tau=\pm 1$)
are shifted up and down by $\sqrt{2}\gamma$ (see Figure
\ref{Fig.energy}(a)). We notice that the existence of three Dirac
points
located at $E =V+\tau \sqrt{2}\gamma$, with $\tau=0, \pm 1$, and two
among them
($\tau=\pm 1$) are shifted either by $\sqrt{2}\gamma$
or by $-\sqrt{2}\gamma$.
In addition, by increasing the potential height $V$, the energy
bands are shifted upwards also by $V$.
To see the effect of the interlayer potential difference $\delta$,
we plot the energy bands for $\delta=2 \gamma$ and $V=0$ in
Figure \ref{Fig.energy}(b). When $\delta\neq 0$, one of the two Dirac
points ($\tau=\pm 1$) is shifted by
$\gamma^{'}=\sqrt{\delta^2+\left(\sqrt{2}\gamma\right)^2}$ and the
other by $-\gamma^{'}$, while the third one
($\tau=0$) remained in the same position.
The effect of $\delta$ can also be taken into account by a
renormalization of the interlayer hopping energy of AAA-stacked
trilayer graphene, $\sqrt{2}\gamma$, to a new interlayer potential
difference dependent hopping energy, $\gamma^{'}$.

\begin{figure}[h!]
 \centering
 \includegraphics[width=7.8cm, height=2.5cm]{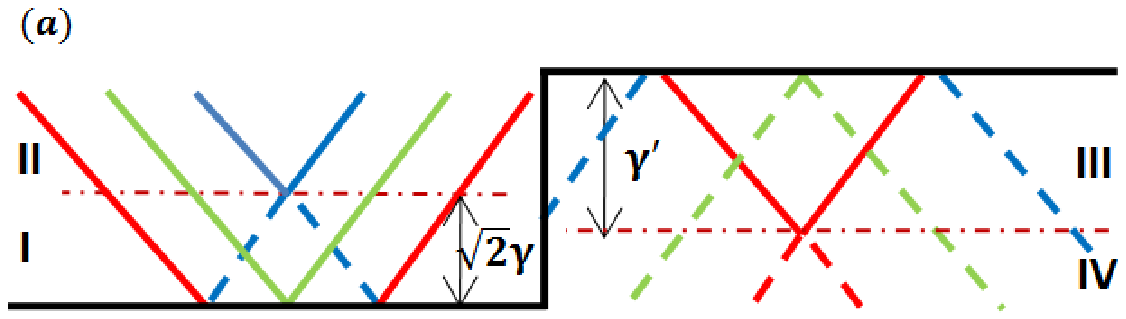}
\ \ \
 \includegraphics[width=7.8cm, height=2.5cm]{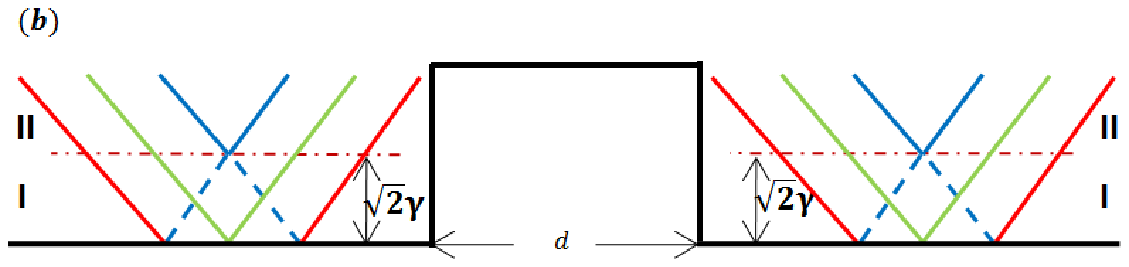}
 \caption{\sf Schematic representation of the band structures of electrons (holes) outside
(inside) the barrier regions. (a): for the {\it np} junction with finite
gap.
 (b): for the barriers structure or {\it npn} junction.}\label{Fig.bandstructure}
\end{figure}

Figure \ref{Fig.bandstructure} presents the band structure of
AAA-stacked trilayer for single and double gates. In the incident
region, the electron states can be subdivided into two regimes.
The first one (\textbf{I}) ($E < \sqrt{2}\gamma $) where
$\tau=+1$ the electron is hole-like particle, while where $\tau=0,-1$ the
electrons are electron-like particle. The second one (\textbf{II}) ($E
> \sqrt{2}\gamma $) where for $\tau=0, \pm 1$ the electrons are
electron-like. However, inside the barrier region, the holes can
also have two chiralities as denoted by regimes (\textbf{III}) and
(\textbf{IV}).
Note that we have nine transmissions that are related to the band
structures in the {\it np} or {\it npn} junctions. In addition, we notice that
for the intra-cone transitions (\textit{i.e.} $\tau \longrightarrow
\tau$ processes), there exist four cases for transitions across
the {\it np} junction (Figure \ref{Fig.bandstructure} (a)):
(1) electron in regime \textbf{I} $\longrightarrow$ hole in regime
\textbf{III}, (2) electron in regime \textbf{I} $\longrightarrow$
hole in regime \textbf{IV}, (3) electron in regime \textbf{II}
$\longrightarrow$ hole in regime \textbf{III}, and (4) electron in
regime \textbf{II} $\longrightarrow$ hole in regime \textbf{IV}.
For the {\it npn} junction (Figure \ref{Fig.bandstructure} (b)):
(1) electron in regime \textbf{I} $\longrightarrow$ electron in
regime \textbf{I}, (2) electron in regime \textbf{I}
$\longrightarrow$ electron in regime \textbf{II}, (3) electron in
regime \textbf{II} $\longrightarrow$ electron in regime
\textbf{I}, and (4) electron in regime \textbf{II}
$\longrightarrow$ electron in regime \textbf{II}.
However, all inter-cone transitions (\textit{i.e.} $\tau
\rightarrow -\tau$ processes) are strictly forbidden due to the
orthogonality of electron wave functions with a different cone
indexes \cite{Sanderson}. As already mentioned above, the
transitions depend on the incident and transmission regions, so
they depend on the presence or absence of the band gap
induced by the single or double gates. For each of the four transitions cases
the transmission can be calculated using the same
method. 
The cone index $\tau$ is introduced while
the chirality $s$ of the massless Dirac quasiparticles is not
necessarily conserved during a transition state \cite{Sanderson}.

\section{Transmission probabilities}

Next we shall show in detail the calculation of the transmission
probabilities. The transfer-matrix method together with
appropriate boundary conditions was implemented for electrons
across the {\it np} and {\it npn} junctions in our
system. We notice that for AAA-stacked trilayer graphene, in the
six band model, we have six reflection and six transmission
channels \cite{Duppen}. Since, for AAA-stacked trilayer graphene,
all inter-cone transitions ($\tau \longrightarrow -\tau$) are strictly
forbidden, then only three transmissions ($\tau
\longrightarrow \tau$) are possible, one of monolayer-like and the two others of
bilayer-like. Thus, we can reduce the six-band model to the
following two-band model of monolayer-like ($\tau=0$) and
four-band model of AA-stacked bilayer-like ($\tau=\pm 1$). The
plane-wave solutions for the Schr\"{o}dinger equation can be
represented by
\begin{equation}
\psi_{j}^{0}=L_{j}^0 \cdot A_{j}^{0}, \qquad \psi_{j}^{\pm
1}=L_{j}\cdot A_{j}
\end{equation}
where $L_{j}^0$ and  $L_{j}$  are
defined by
\begin{eqnarray}\label{L}
{L_{j}^0= \left(%
\begin{array}{cc}
  s_j e^{-i\phi_{j}^0}e^{ik_{x_{j}}^0 x} & -s_j e^{i\phi_{j}^0}e^{-ik_{x_{j}}^0
  x}\\
  e^{ik_{x_{j}}^0 x} &  e^{-ik_{x_{j}}^0 x} \\
  \end{array}%
\right)}
\end{eqnarray}
\begin{eqnarray}
{
L_{j}=\left(%
\begin{array}{cccc}
  s_j e^{-i\phi_{j}^{+}}e^{ik_{x_{j}}^+ x} & -s_j
  e^{i\phi_{j}^+}e^{-ik_{x_{j}}^+
  x} & s_j e^{-i\phi_{j}^{-}}e^{ik_{x_{j}}^- x} & -s_j
  e^{i\phi_{j}^-}e^{-ik_{x_{j}}^-
  x}\\
  e^{ik_{x_{j}}^+ x} &  e^{-ik_{x_{j}}^+ x} & e^{ik_{x_{j}}^- x} &  e^{-ik_{x_{j}}^- x} \\
   \frac{s_j\sqrt{2}}{\sqrt{\delta^2+2}}e^{-i\phi_{j}^{+}}e^{ik_{x_{j}}^+ x} & -\frac{s_j\sqrt{2}}{\sqrt{\delta^2+2}}
  e^{i\phi_{j}^+}e^{-ik_{x_{j}}^+
  x}&  -\frac{s_j\sqrt{2}}{\sqrt{\delta^2+2}}e^{-i\phi_{j}^{-}}e^{ik_{x_{j}}^- x} & \frac{s_j\sqrt{2}}{\sqrt{\delta^2+2}}
  e^{i\phi_{j}^-}e^{-ik_{x_{j}}^-
  x}\\
 \frac{\sqrt{2}}{\sqrt{\delta^2+2}} e^{ik_{x_{j}}^+ x} &
 \frac{\sqrt{2}}{\sqrt{\delta^2+2}}e^{-ik_{x_{j}}^+ x}& -\frac{\sqrt{2}}{\sqrt{\delta^2+2}} e^{ik_{x_{j}}^- x} &  -\frac{\sqrt{2}}{\sqrt{\delta^2+2}}e^{-ik_{x_{j}}^- x} \\
  \end{array}%
\right).}\lb{LL}
\end{eqnarray}
The index $j$ denotes each potential region, $j= 0$ for the
incident region, $j = 1$ for the potential barrier region and $j =
2$ for the transmission region,
$\phi_{j}^{\tau}=\arctan{(k_y/k_{x_{j}}^\tau)}$ are the phase,
$A_{j}^0$ and $A_{j}$ are defined by $A_{1}^0=(\alpha_0,
\beta_0)^T$ and $A_{1}=(\alpha_+, \beta_+, \alpha_-, \beta_-)^T$.
We are interested in the normalization coefficients, the
components of $A_{0}^0$ and $A_{0}$, on both sides of the {\it np}
and {\it npn} junction structure. In other words in the incident
region we have $A_{0}^0 = (1, r_0 )^T$ and $A_{0} = (1, r_+,1,
r_-)^T$ and in the transmission region we have $A_{2}^0 = (t_0,0
)^T$ and $A_2 = (t_+, 0,t_-, 0)^T$, where $r_\tau$ and $t_\tau$
are the reflection and transmission coefficients of each cone
($\tau=0,\pm 1$), respectively.

We need to match the wave functions at the boundaries between
different regions. This procedure is most conveniently expressed
in the transfer matrix formalism \cite{Wang}. After a
straightforward algebra we get the transfer matrix
for {\it np} and {\it npn} junctions
\begin{eqnarray}\label{M}
&& M_{{np}}^0=(L_{0}^0)^{-1}[x=0]\cdot L_{1}^0[x=0]\\
&&  M_{{npn}}^0=(L_{0}^0)^{-1}[x=0]\cdot
L_{1}^0[x=0]\cdot (L_{2}^0)^{-1}[x=d]\cdot L_{2}^0[x=d]
\\
&&
M_{{np}}=L_{0}^{-1}[x=0]\cdot L_{1}[x=0]\\
&&
M_{{npn}}=L_{0}^{-1}[x=0]\cdot L_{1}[x=0]\cdot
L_{1}^{-1}[x=d]\cdot L_{2}[x=d].
\end{eqnarray}
Here $L_{0}=L_{2}$ is determined by \eqref{L} and \eqref{LL} at $V = \delta = 0$.
Consequently, we have three channels for the transmission
probability in each individual cone ($\tau=0,\pm 1$) for {\it np} and
{\it npn} junctions.

\subsection{{\it np} junction}

The transmission probabilities for the intracone transitions with
different cone index (\textit{i.e.} $\tau \longrightarrow -\tau$
processes) is zero \cite{Sanderson}. This can be understood by
considering the step potential as a sharp perturbation. While for
the intercone transitions with the same cone index (\textit{i.e.}
$\tau \longrightarrow \tau$ processes) we have three transitions given by
\begin{eqnarray}
&& T_{{np}}^+=\frac{1}{\left(M_{{np}}[1,1]\right)^2}\\
&& T_{{np}}^-=\frac{1}{\left(M_{{np}}[3,3]\right)^2}\\
&& T_{{np}}^0=\frac{1}{\left(M_{{np}}^0[1,1]\right)^2}
\end{eqnarray}
where $M_{{np}}[1,1]$ and $M_{{np}}[3,3]$ are the
elements of the transfer matrix $M_{{np}}$ given above. 

\begin{figure}[h!]
 \centering
\includegraphics[width=5.5cm, height=5cm]{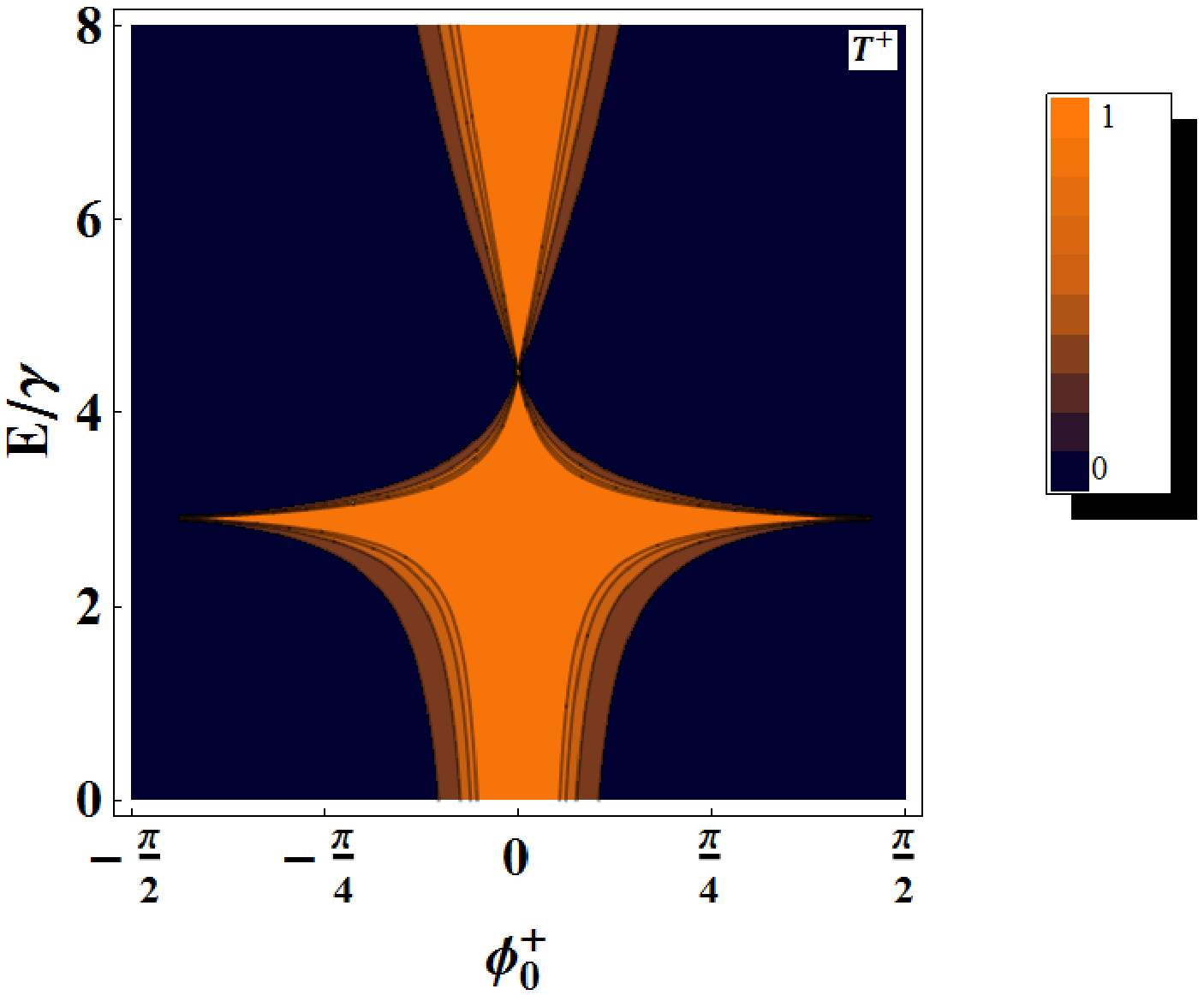} \
 \includegraphics[width=5.5cm, height=5cm]{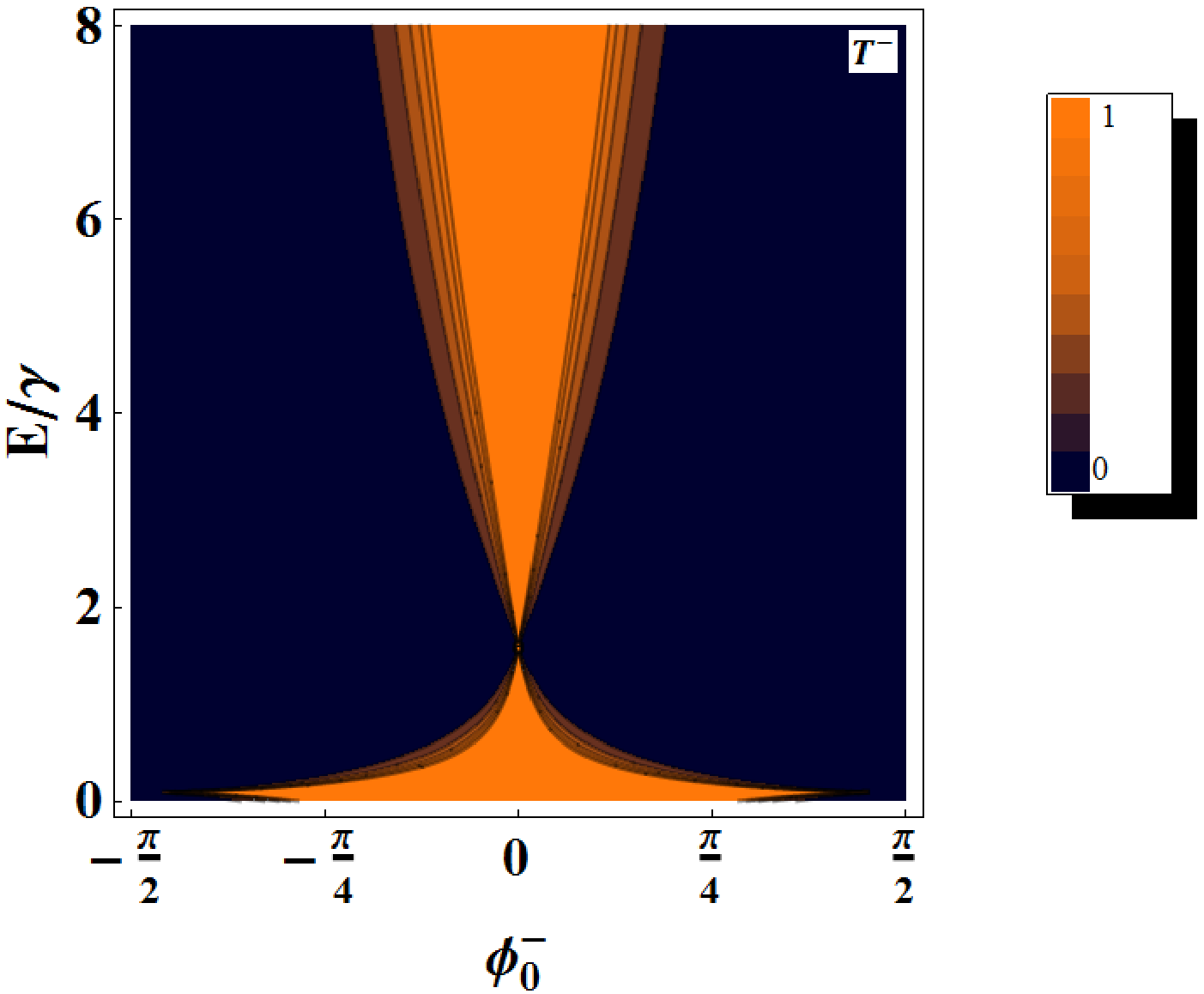}
 \
  \includegraphics[width=5.5cm, height=5cm]{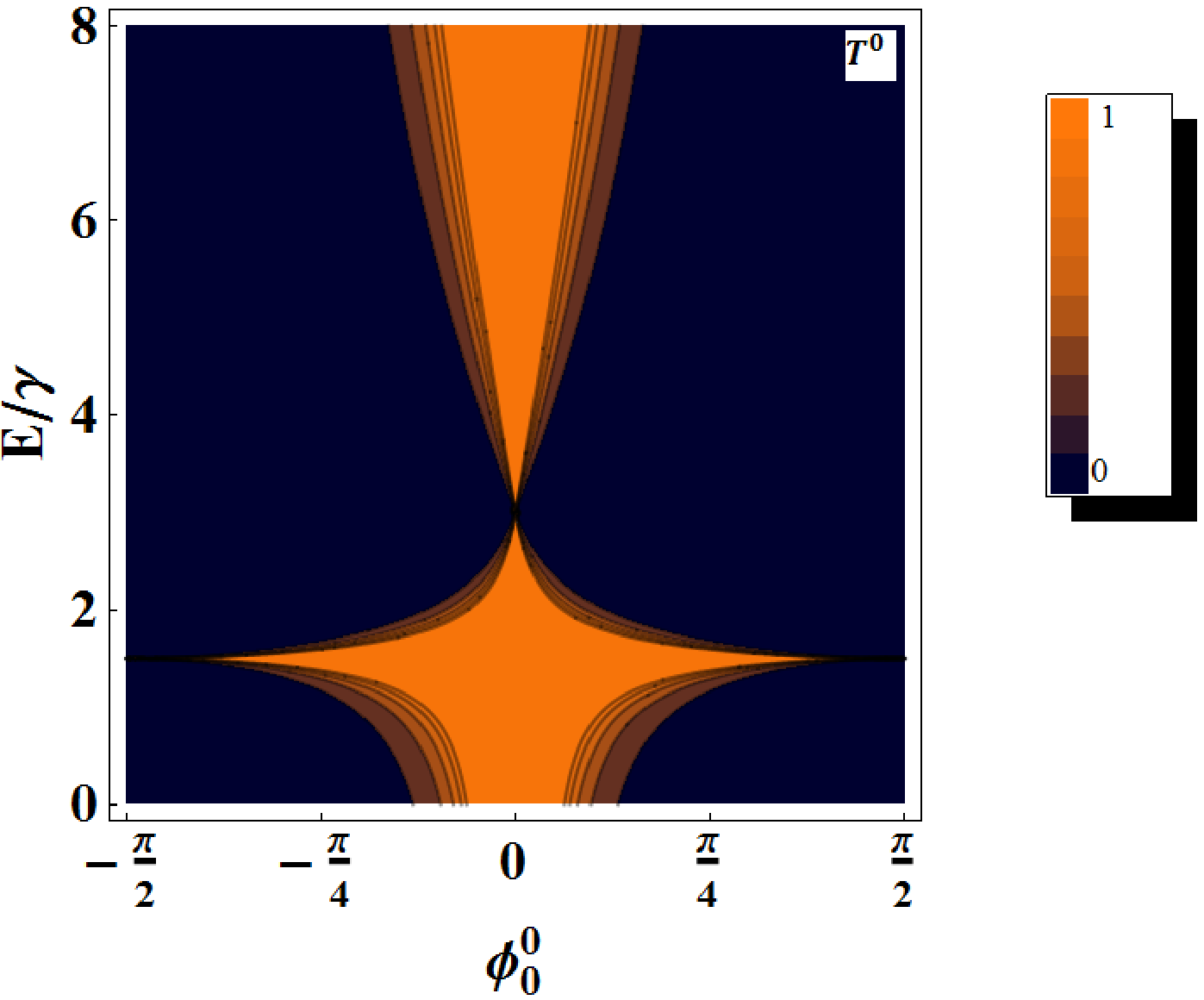}
 \caption{\sf Contour plot of the three transmission probabilities  as function of the incident
angle for {\it np} junction, with physical parameters: $V =3  \gamma
$ and $\delta=0$.}\label{contour-T}
\end{figure}

We first consider the case of {\it np} junction with $\delta=0$. In
Figure \ref{contour-T}, we present the contour plot of the three
transmissions: one of monolayer-like ($T^0$) and two of AA-stacked
bilayer-like ($T^\pm$), as a function of the incident angle and
its energy, with physical parameters $V = 3 \gamma$ and
$\delta=0$. We can clearly see that the three transmission
($T^\tau$) exhibits a maximum (perfect transmission) for
$E=\frac{V}{2}+\tau\sqrt{2}\gamma$ \cite{Sanderson} and for normal
incidence $\phi^\tau\longrightarrow 0$, as predicted \cite{Katsnelson}
and observed experimentally \cite{Young,Stander}. The major
difference between the monolayer-like ($T^0$) and
 AA-stacked bilayer-like ($T^\pm$) is that the monolayer-like is only described by
 pseudospin, while the bilayer-like is described by an additional cone
 index, {\it i.e.} $\tau$. In addition, the chirality of electron in the incident region
($\tau=0, \pm$)  is always $s_0=+1$, while for the hole in transmission region it
can be $ s = \pm 1$. For $s=+1$ we have the usual
refraction and for $s=-1$ the electron transmission is equivalent
to the optical case of negative refraction. We recall that for
AA-stacked bilayer-like ($T^\pm$) the band structure is composed
of two Dirac cones shifted by $\tau\sqrt{2}\gamma$, one can clearly
see that the transmission for both cones ($T^\pm$), has the same
form as that in the case of monolayer-like ($T^0$). We observe
also that the three transmissions curves ($T^\tau$) are
symmetrical with respect to the normal incidence and the three
Dirac points are located at $E=V+\tau\sqrt{2}\gamma$, with $\tau=0, \pm
1$.

\subsection{{\it npn} junction}

Now, we use
\eqref{M} to explore the electronic transport properties through an
{\it npn} junction
based on the AAA-stacked
trilayer graphene. The three transmissions are defined by
\begin{eqnarray}
&& T_{{npn}}^+=\frac{1}{\left(M_{{npn}}[1,1]\right)^2}\\
&& T_{{npn}}^-=\frac{1}{\left(M_{{npn}}[3,3]\right)^2}\\
&& T_{{npn}}^0=\frac{1}{\left(M_{{npn}}^0 [1,1]\right)^2}.
\end{eqnarray}
\begin{figure}[h!]
 \centering
\includegraphics[width=5.5cm, height=5cm]{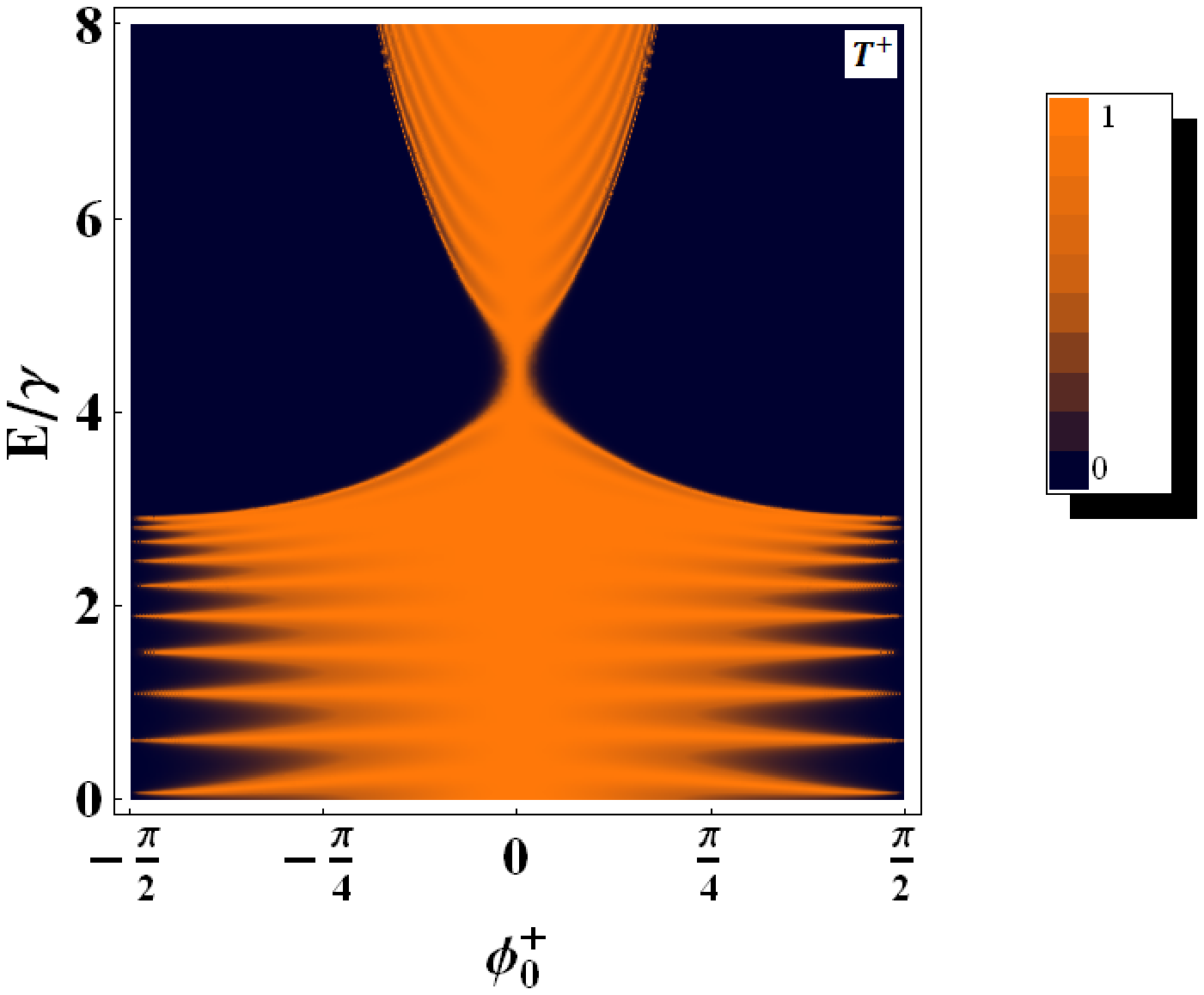}
 \
\includegraphics[width=5.5cm, height=5cm]{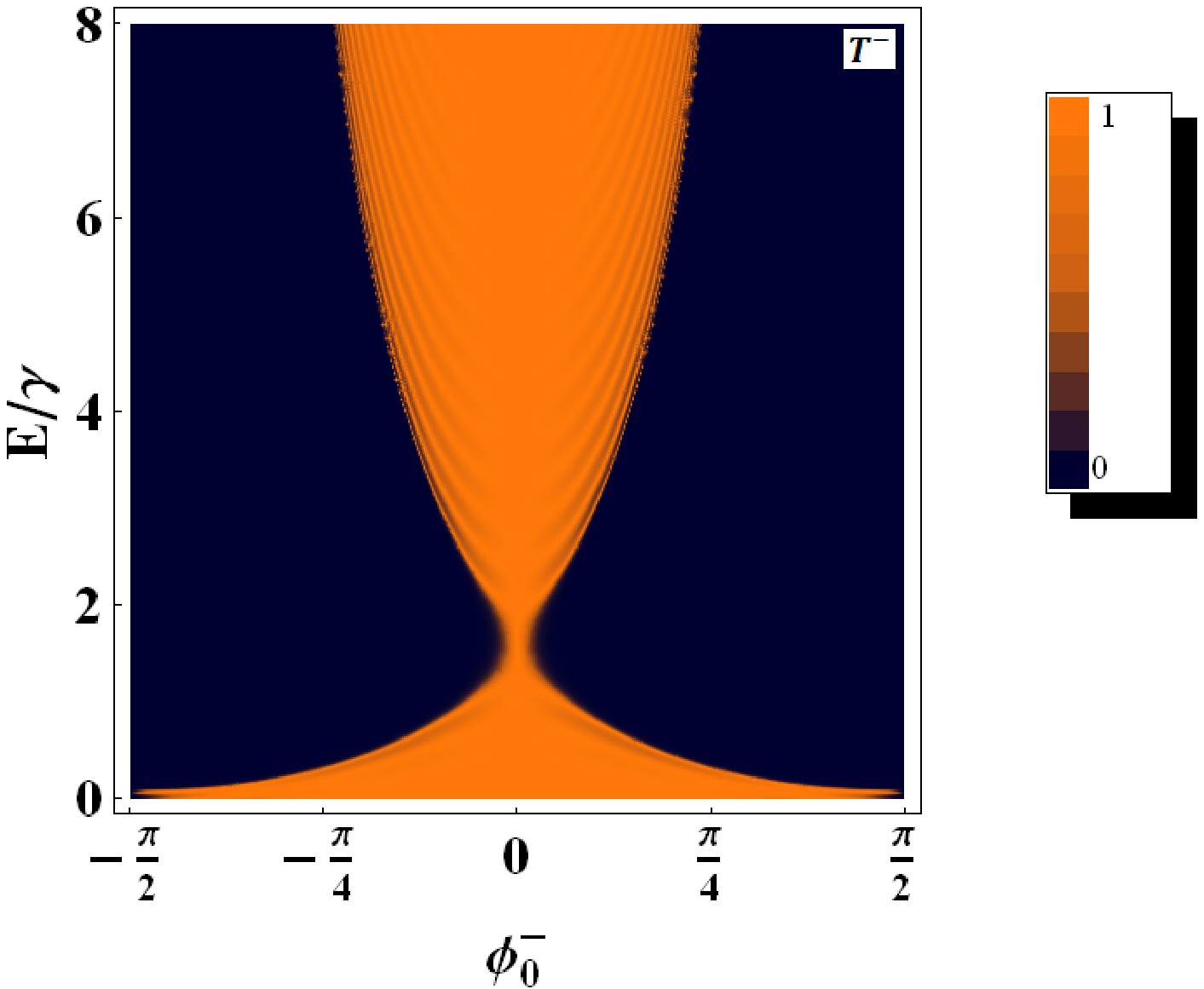}
  \
 \includegraphics[width=5.5cm, height=5cm]{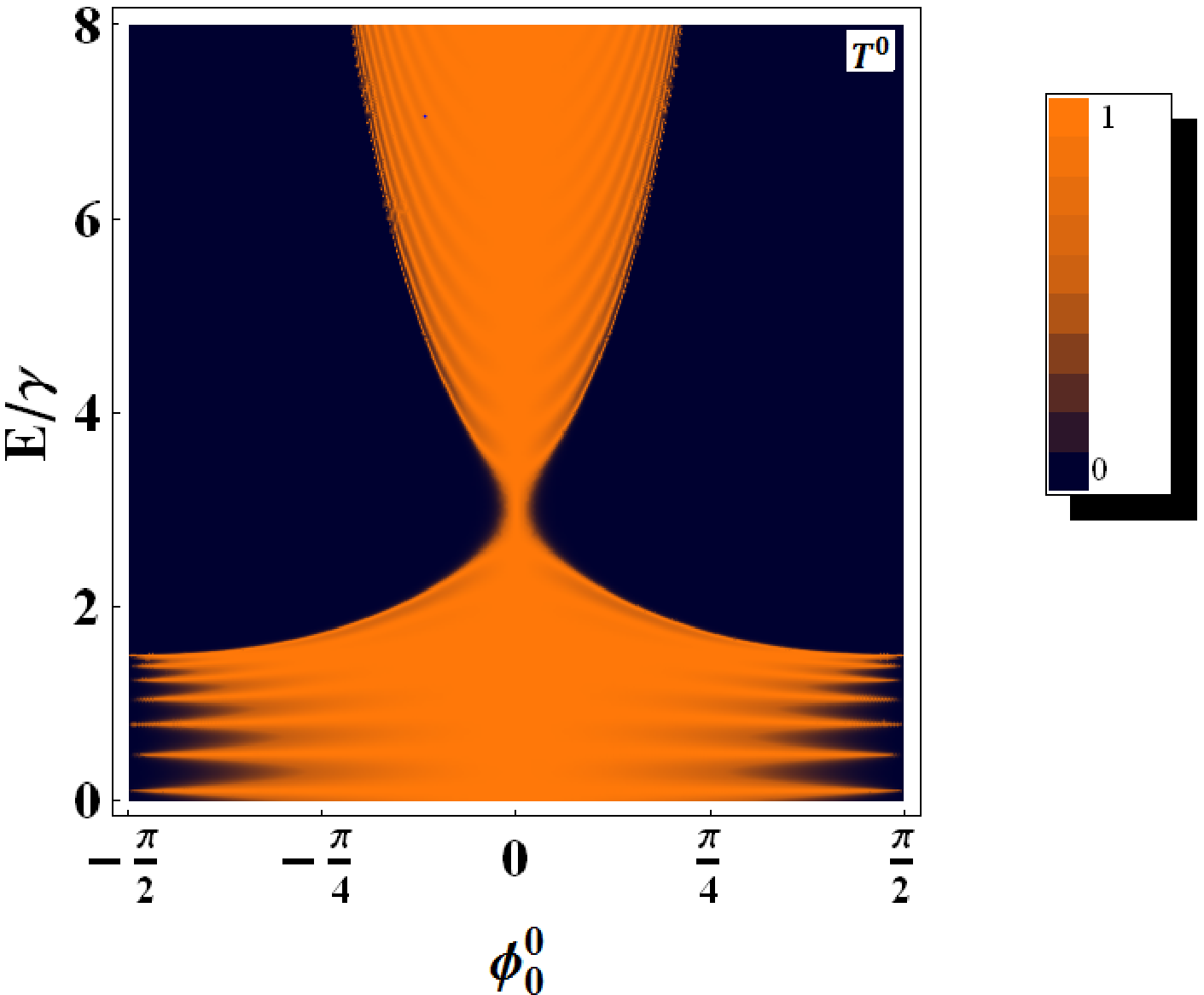}
 \caption{\sf Density plot of the three transmission probabilities as a function of the incident angle and its energy
for a single barrier with a single gate ($\delta=0$), with physical
parameters: $ d=25  nm$, $V =3  \gamma $, and
$\delta=0$.}\label{Density-T}
\end{figure}
In the following calculation, we start with an {\it npn} junction in the
absence of double gates. In Figure \ref{Density-T}, we show the
density plot of the three transmissions ($T^\tau$) as a function
of the incident angle ($\phi^\tau$) and its energy for both cones
with physical parameters: $ d=25  nm$, $V =3 \gamma $, and
$\delta=0$. The different colors from blue to orange correspond to
different values of the transmission from 0 to 1. It is important
to note that in the case of AAA-stacked trilayer graphene the band
structure is composed of three copies of the monolayer band
structure \cite{Chen}. One of them ($\tau=1$) is shifted
by $\sqrt{2}\gamma$ and the other ($\tau=-1$) by $-\sqrt{2}\gamma$,
while the third one ($\tau=0$) remains in the center.
Then we end up with three Dirac points located at $E=V+\tau
\sqrt{2}\gamma$, with $\tau= 0,\ \pm 1$. For the transmission
$T^0$ of the center layer the Dirac points correspond to $E=V$,
for the top layer $T^+$ correspond to $E=V+\sqrt{2}\gamma$ and the
bottom layer $T^-$ correspond to $E=V-\sqrt{2}\gamma$.
In the case of AA-stacked bilayer-like, for both cones
($\tau=\pm 1$), it has the same form as that in the case of
monolayer-like ($\tau=0$). 
However, for AA-stacked bilayer-like graphene both the electrons
and holes have the same chirality index $s = \pm 1$ while in the
case of monolayer-like graphene the electron has always $s=+1$ and
$s=-1$ for the hole. We can see from $T^\tau$ that there is
a perfect transmission for normal or near normal incidence
($\phi^\tau\longrightarrow 0$), which is a manifestation of the Klein
tunneling \cite{Katsnelson}. These results were also found in the
case of monolayer graphene and AA-stacked bilayer graphene.
Moreover, we find that the single gate ($\delta=0$) behavior in
AAA-stacked trilayer graphene is a superposition of monolayer-like
and bilayer-like systems, which are similar to those
obtained for ABA-stacked trilayer graphene \cite{Duppen}.\\

\begin{figure}[h!]
 \centering
 \includegraphics[width=5.5cm, height=5cm]{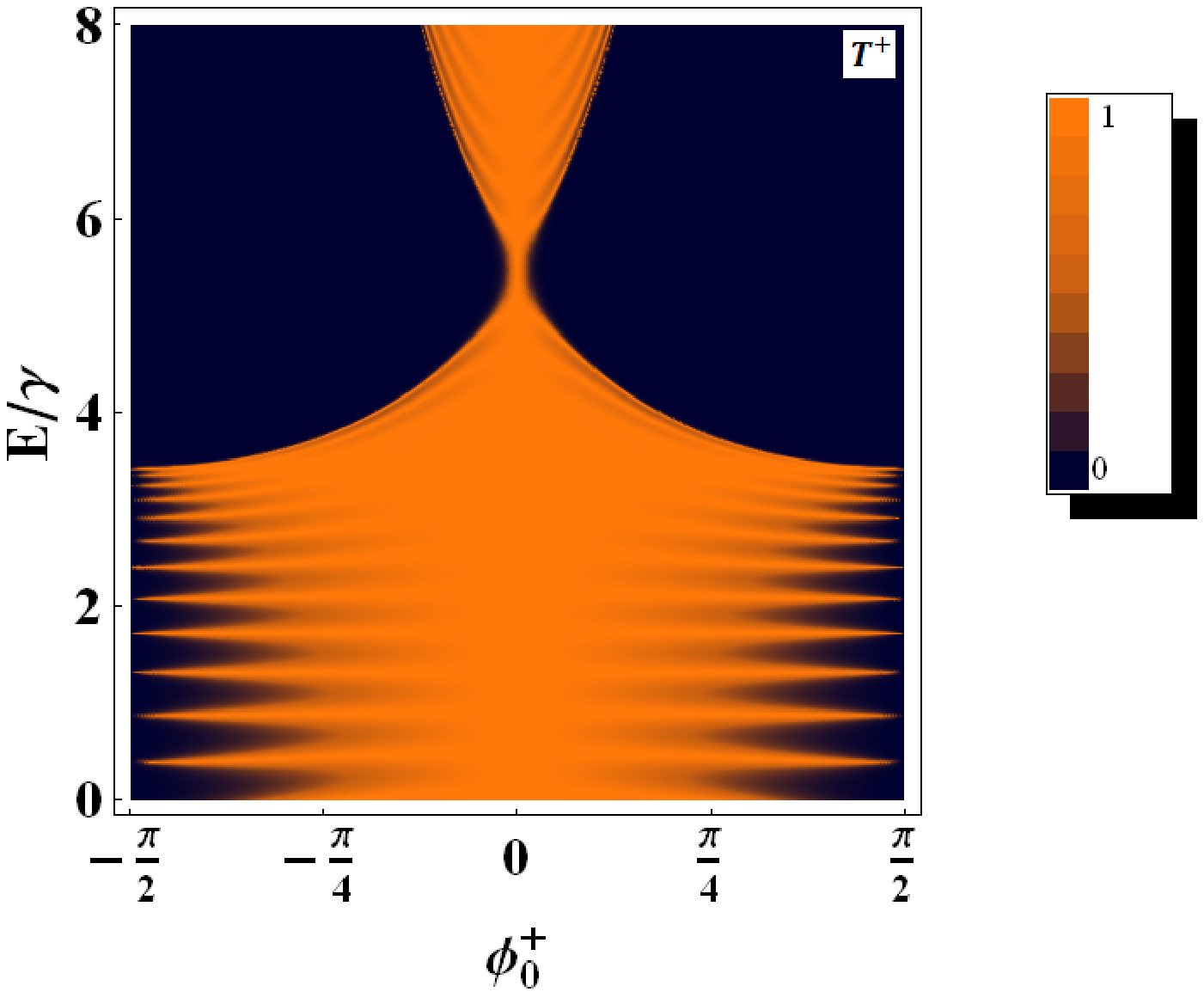}
 \
  \includegraphics[width=5.5cm, height=5cm]{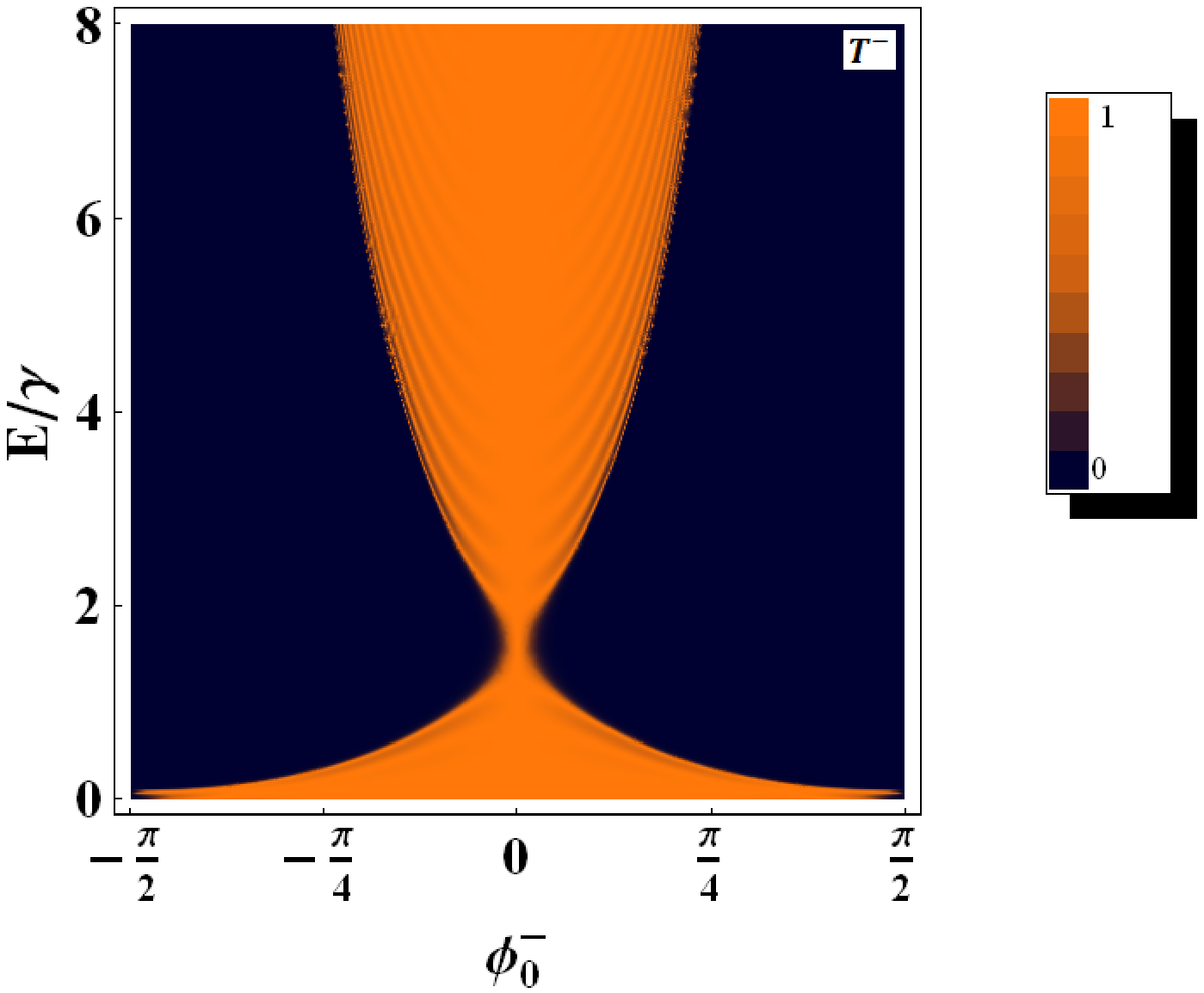}
    \
 \includegraphics[width=5.5cm, height=5cm]{D-T0}
 \caption{\sf Density plot of the three transmission probabilities in AA-stacked bilayer-like as a function of the incident angle and its energy
for single barrier with double gate ($\delta=2 \gamma$), and
physical parameters: $ d=25  nm$ and $V =3  \gamma
$.}\label{Fig.DnsityPlotT-delta}
\end{figure}

Now, we will turn to the discussions of the influence of the
interlayer potential difference $\delta$ on the three
transmissions $T^\tau$.
We show the density plot of the three transmissions as function of
the incident angle and its energy in Figure
\ref{Fig.DnsityPlotT-delta} with $V=3 \gamma$, $d=25  nm$ and
$\delta=2 \gamma$. We notice that two Dirac points
($\tau=\pm 1$) are shifted by
$\gamma^{'}=\sqrt{\delta^2+\left(\sqrt{2}\gamma\right)^2}$ and the
other by $-\gamma^{'}$ and the third one
$\tau=0$ remains in the same position (see Figure \ref{Density-T}).
These cases do not exist in the AA-stacked bilayer graphene
\cite{Sanderson}, but in the presence of  $\delta$ we have a
transmission gap around the Dirac point. As already mentioned
above,
we have perfect transmission, which is a manifestation of Klein
tunneling effect. However, in the case of AAA-stacked trilayer
graphene, the effects of $\delta$ can be taken into account by a
renormalization of the interlayer hopping energy,
$\sqrt{2}\gamma$, to a new interlayer potential difference
dependent hopping energy,
$\gamma^{'}=\sqrt{\delta^2+\left(\sqrt{2}\gamma\right)^2}$. We
notice that in the case AA-stacked bilayer graphene\cite{Sanderson},
for $\delta \neq 0$, the Klein tunneling effect is suppressed.
Also, we find
that the single and double gate behavior in the case of
AAA-stacked trilayer graphene is a superposition of monolayer-like
and bilayer-like systems. This is not the case for ABA-stacked
trilayer graphene, where the double gate mixes both types of bands
and breaks the angular symmetry with respect to normal incidence
\cite{Duppen}.

\section{Conductance}

\begin{figure}[h!]
 \centering
 \includegraphics[width=6.5cm, height=5cm]{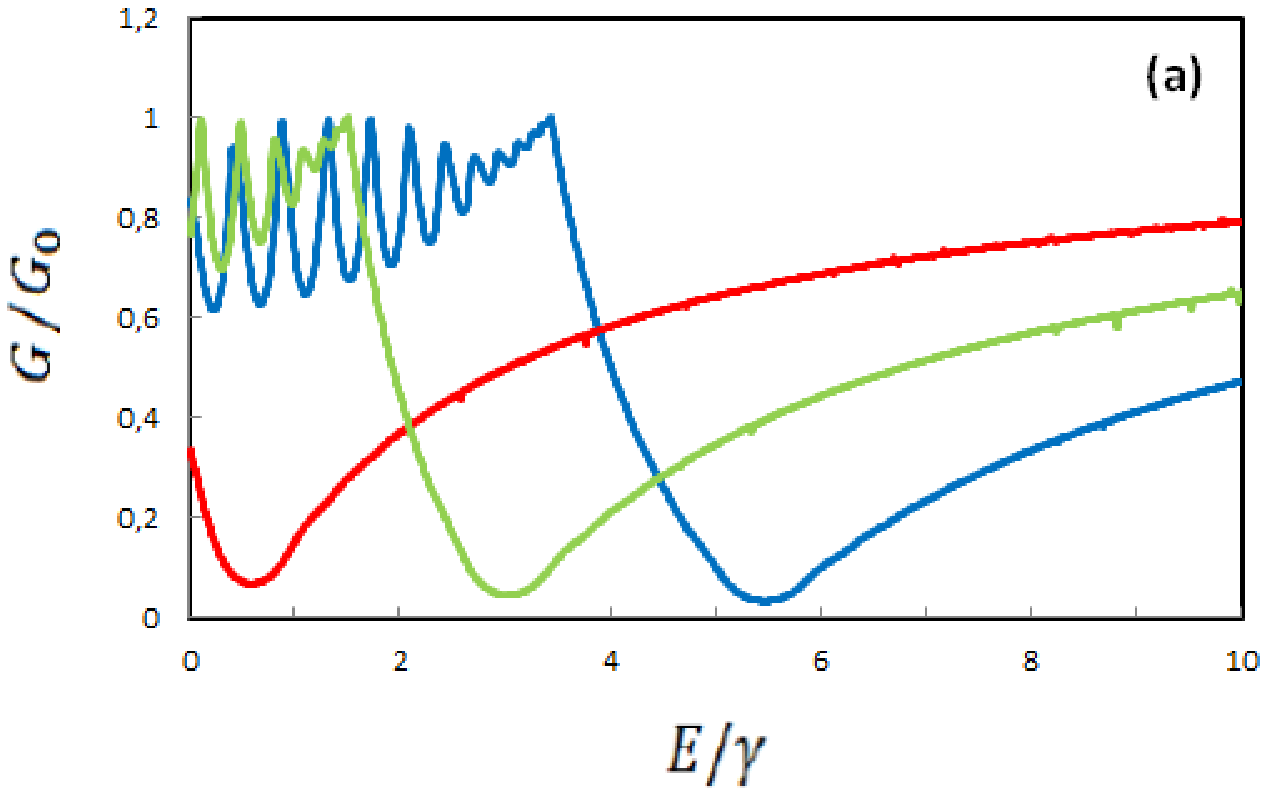}
 \ \ \ \
  \includegraphics[width=6.5cm, height=5cm]{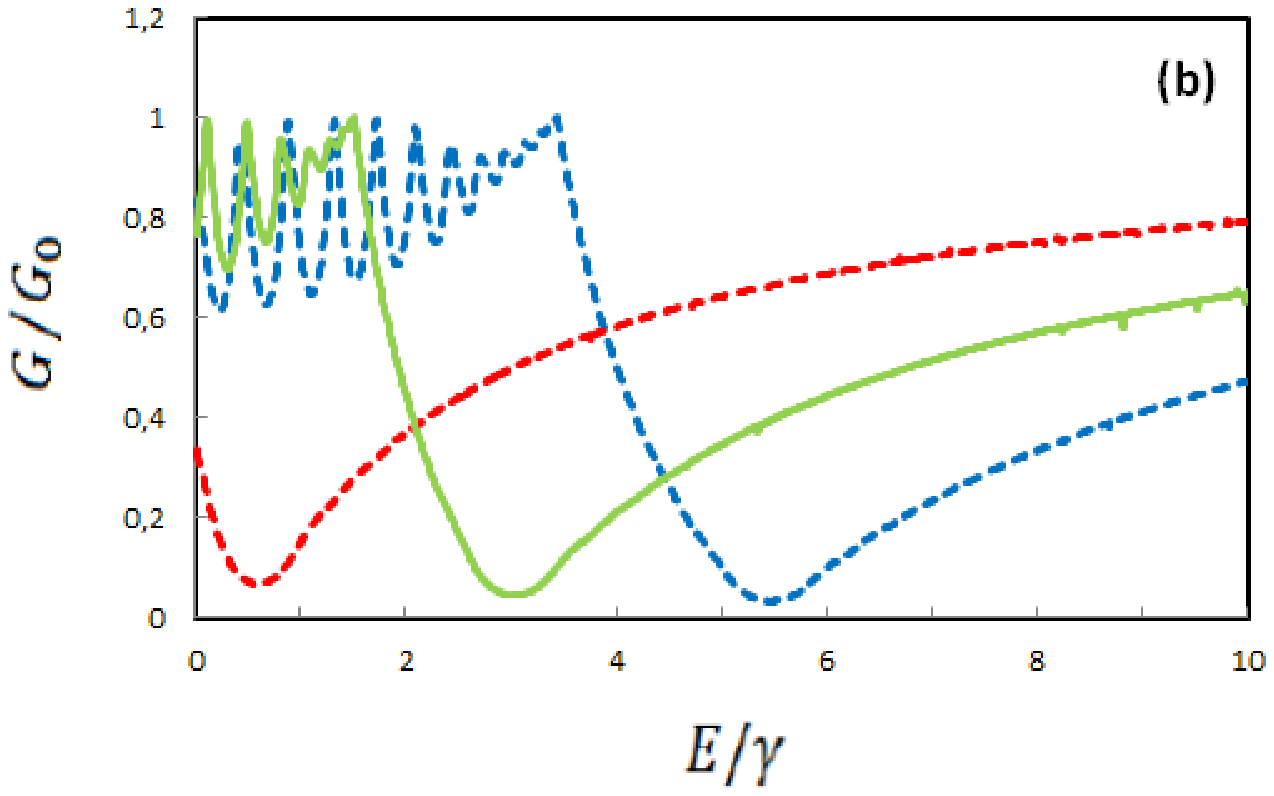}
     \caption{\sf Conductances in each individual  cone ($\tau=0, \pm 1$) through $npn$ junction as a function of the energy, with $V=3\gamma$  and
$d=25nm$. (a): for $\delta=0$, $\tau=-1$ (red line), $\tau=0$
(green line) and $\tau=+1$ (blue line). (b): for $\delta=2\gamma$,
$\tau=-1$ (red dashed), $\tau=0$ (green line) and $\tau=+1$ (blue
dashed).}\label{Fig.conductance}
\end{figure}

The conductance through $npn$ junction can be expressed in terms
of the transmission probabilities established before. We will see how the three
conductance $G^\tau$ of each cone channel ($\tau =0, \pm 1$)
will behave. For this purpose, we evaluate $G^\tau$ by using the
Landauer-B\"{u}ttiker formula \cite{Blanter}
\begin{equation}
G^\tau=G_0\int_{0}^{\pi/2} T^\tau
\left(E,\phi_{0}^\tau\right)\cos{\phi_{0}^\tau}d\phi_{0}^\tau
\end{equation}
where the unit conductance is given by
\beq G_0=N L_y k_F e^2/h\pi 
\eeq 
and the
factor $N=4$ is due to the spin and valley degeneracy,  $L_y$ is
the width of the sample in the $y$-direction and 
\beq
k_F=\sqrt{k_{y}^2+(k_{x_0}^\tau)^2}=s_0(E-\tau\sqrt{2}). 
\eeq 
The
total conductance $G_t$ is defined as the sum of the conductance
channels in each individual cone $G^\tau$ such as
\begin{equation}\label{condtotal}
G_t=\frac{1}{3}\left(G^+ +G^-+G^0\right)
\end{equation}
where the factor $1/3$ is required since the total conductance is
a contribution of three cones. Whereas, in the case of AA-stacked
bilayer graphene, there are two transmissions channels then the
total conductance is the sum of two conductances channels with a
factor of $1/2$ \cite{Sanderson}. In the forthcoming analysis, we
evaluate numerically the conductance in AAA-stacked bilayer
graphene. In Figure \ref{Fig.conductance} we show the three
conductance through a single barrier structure ($npn$ junction) in
each individual cone as a function of the energy for $V=3\gamma$
and $d=25nm$. In Figure \ref{Fig.conductance}(a) the conductances
are plotted for $\delta=0$ while in Figure
\ref{Fig.conductance}(b) for $\delta=2\gamma$. It is important to
note that in the case of AAA-stacked trilayer graphene, in Figure
\ref{Fig.conductance}(a), we have the three conductances for each
individual cone ($\tau$).
 One of them ($\tau =+1$) is shifted by $\sqrt{2}$ and the other
($\tau =-1$) by $-\sqrt{2}$, while the third one ($\tau =0$)
remains in the center. Then we end up with three Dirac points
located at $E = V + \tau \sqrt{2}$, with ($\tau=0,\pm 1$). For
energies smaller than $V+\tau\sqrt{2}$, the conductance of the
single barrier presents  peaks.
While for $E>V+\tau\sqrt{2}$, the conductance increases and the
peaks are absents.  To see the effect of the externally induced
interlayer potential difference $\delta$, we plot the three
conductances as function of the energy in Figures
\ref{Fig.conductance}(b). We note that two Dirac points for
$G^{\tau=+ 1}$ and $G^{\tau=- 1}$ are shifted by
$\gamma^{'}=\sqrt{\delta^2+\left(\sqrt{2}\gamma\right)^2}$ and the
other by $-\gamma^{'}$. Moreover, the third Dirac point for
$G^{\tau=0}$ remains in the same position.\\

\section{Conclusion}

We have investigated the electronic transport
properties through {\it np} and {\it npn} junctions based on the AAA-stacked
trilayer graphene.
The Hamiltonian model describing the system under consideration
allowed us to determine the solutions of the energy spectrum.
The obtained bands are composed of three Dirac cones that depend
on the cone and chirality indexes. For AAA-stacked
trilayer graphene, we have shown that the transmission in the inter-cone transition
($\tau\longrightarrow -\tau$ processes) is strictly forbidden due to
the orthogonality of electron wave functions, in contrast of the
ABA- and ABC-stacked trilayer graphene cases.
In addition, our results showed that there are three transmission
probabilities for each individual cone ($\tau=0, \pm 1$), one of
monolayer-like $T^0$ and two of AA-stacked bilayer-like $T^\pm$.
The quasiparticles in AAA-stacked trilayer graphene are not only
chiral but also labeled by an additional cone index $\tau$.
It was noticed that $\tau$ is a strictly conserved quantity while the
chirality index $s$ is not necessarily conserved during a state
transition.

Subsequently, we have numerically investigated the obtained three transmission
probabilities for each
individual cone. 
For {\it np} junction with single gate, we have perfect transmission
for $E=\frac{V}{2}+\tau\sqrt{2}\gamma$ and for normal incidence
$\phi^\tau\longrightarrow 0$. In addition, we have found that the three
transmissions $T^\tau$ have the same form as those found in the case
of monolayer graphene. We have shown that there are three Dirac points
located at $E=V+\tau\sqrt{2}\gamma$ where $\tau =0, \pm 1$. Also,
we have found that the single and double gated behavior are a superposition
of monolayer-like and bilayer-like systems. These results are
similar to ABA-stacked trilayer graphene for a single gate.
However, double gated mixes both types of bands in the case of
ABA-stacked trilayer graphene. Furthermore, we have studied the
influence of the double gated on the three transmission $T^\tau$
for single barrier structure and  shown that it can be taken into
account by a renormalization of the interlayer hopping energy to a
new interlayer potential
$\sqrt{2}\gamma \longrightarrow
 \gamma^{'}=\sqrt{\delta^2+\left(\sqrt{2}\gamma\right)^2}$.

{Based on the obtained results for
the transmission probabilities we have found  three
conductances channels through $npn$ junction.
For $\delta=0$, that there are three Dirac points located at
$E=V+\tau\sqrt{2}$ with $\tau=0, \pm 1$. Also, we have
investigated the effect of the interlayer potential difference
$\delta$ on the conductance channel in each cone. It was noticed
that the two Dirac point for $G^{\tau=+ 1}$ and $G^{\tau=- 1}$ are
shifted by
$\gamma^{'}=\sqrt{\delta^2+\left(\sqrt{2}\gamma\right)^2}$ and the
other by $-\gamma^{'}$. Moreover, the third Dirac point for
$G^{\tau=0}$ remains in the same position. Finally the total
conductance is analyzed  as the average of the three conductances
channels in each cone.}

\section*{Acknowledgments}

The generous support provided by the Saudi Center for Theoretical
Physics (SCTP) is highly appreciated by all authors. {AH and HB
acknowledge the support of King Fahd University of Petroleum and
minerals under research group project RG1502-1 and Rg1502-2.}


\begin{thebibliography}{99}


\bibitem{Novoselov}          K. S. Novoselov, A. K. Geim, S. V. Morozov, D. Jiang, Y. Zhang, S. V. Dubonos,
I. V. Grigorieva, and A. A. Firsov, Science 306, 666 (2004).
\bibitem{Novoselov2}         K. S. Novoselov, A. K. Geim, S. V. Morozov, D. Jiang, M. I. Katsnelson,
I. V. Grigorieva, S. V. Dubonos, and A. A. Firsov, Nature 438, 197 (2005).
\bibitem{Zhang}              Y. B. Zhang, Y. W. Tan, H. L. St\"ormer, and P. Kim, Nature 438, 201 (2005).

\bibitem{Berger}             C. Berger {\it et al.}, 
J. Phys. Chem. B 108, 19912 (2004).
\bibitem{Bunch}              J. S. Bunch, Y. Yaish, M. Brink, K. Bolotin, and P. L. McEuen, Nano Lett. 5, 2887 (2005).
\bibitem{Novoselov3}         K. S. Novoselov, D. Jiang, F. Schedin, T. J. Booth, V. V. Khotkevich, S. V. Morozov, and A. K. Geim,
Proc. Nat. Acad. Sc. 102, 10451 (2005).
\bibitem{Novoselov4}         K. S. Novoselov, E. McCann, S. V. Morozov, V. I. Fal'ko, M. I. Katsnelson, U. Zeitler, D. Jiang, F. Schedin,
and A. K. Geim, Nature Physics 2, 177 (2006).
\bibitem{Morozov}            S. V. Morozov, K. S. Novoselov, F. Schedin, D. Jiang, A. A. Firsov, and A. K. Geim, Phys. Rev. B 72, 201401 (2005).

\bibitem{Partoens}             B. Partoens, and F. M. Peeters, Phys. Rev. B 75, 193402 (2007).
\bibitem{Koshino}              M. Koshino, and T. Ando, Phys. Rev. B 77, 115313 (2008).
\bibitem{Nilsson}              J. Nilsson, A. H. Castro Neto, F. Guinea, and N. M. R. Peres, Phys. Rev. B 78, 0454005 (2008).
\bibitem{Avetisyan}            A. A. Avetisyan, B. Partoens, and F. M. Peeters, Phys. Rev. B 80, 195401 (2009).
\bibitem{Koshino2}              M. Koshino, Phys. Rev. B 81, 125304 (2010).
\bibitem{DasSarma}             S. Das Sarma, S. Adam, E. H. Hwang, and E. Rossi, Rev. Mod. Phys. 83, 407 (2011).
\bibitem{Jung}                 J. Jung, F. Zhang, Z. Qiao, and A. H. MacDonald, Phys. Rev. B 84, 075418 (2011).
\bibitem{DeAndres}             P. L. De Andres, F. Guinea, and M. I. Katsnelson, Phys. Rev. B 86, 245409 (2012).
\bibitem{Munoz}                W. A. Munoz, L. Covaci, and F. M. Peeters, Phys. Rev. B 88, 214502 (2013).
\bibitem{Duppen}               B. Van Duppen, S. H. R. Sena, and F. M. Peeters, Phys. Rev. B 87, 195439 (2013).
\bibitem{Duppen2}              B. Van Duppen, and F.M. Peeters, Europhys. Lett. 102, 27001 (2013)
\bibitem{Bala}                 S. Bala Kumar, and J. Guo, Appl. Phys. Lett. 100, 163102 (2012).
\bibitem{Duppen3}              B. Van Duppen, and F. M. Peeters, Appl. Phys. Lett. 101, 226101 (2012).


\bibitem{Quhe12} R. Quhe, J. Zheng, G. Luo, Q. Liu, R. Qin, J. Zhou, D. Yu, S. Nagase,
W.-N. Mei, Z. Gao and J. Lu, 
NPG
Asia Mater. 4, (2012) e6.

\bibitem{Tabert}               C. J. Tabert and E. J. Nicol, Phys. Rev. B 86, 075439 (2012).
\bibitem{Ando2}                T. Ando, J. Phys. Conf. Ser. 302, 012015 (2011).
\bibitem{Hsu}                  Y.-F. Hsu, and G.-Y. Guo, Phys. Rev. B 82, 165404 (2011).
\bibitem{Prada}                E. Prada, P. San-Jose, L. Brey, and H. Fertig, Solid State Commun. 151, 1075 (2011).
\bibitem{Brey}                 L. Brey, and H. A. Fertig, Phys. Rev. B 87, 115411 (2013).
\bibitem{Mohammadi2}           Y. Mohammadi, R. Moradian, Physica B 442, 66 (2014).
\bibitem{Mohammadi}            Y. Mohammadi, R. Moradian, and F. S. Tabar, Solid State Commun. 193, 1 (2014).
\bibitem{Sanderson}            M. Sanderson, Y. S. Ang, and C. Zhang, Physical Review B 88, 245404 (2013).

\bibitem{Wang}                 L.-G. Wang and S.-Y. Zhu, Phys. Rev. B 81, 205444 (2010).
\bibitem{Katsnelson}           M. I. Katsnelson, K. S. Novoselov and A. K. Geim, Nat. Phys. 2, 9 (2006).
\bibitem{Young}                A. F. Young and P. Kim, Nat. Phys. 5, 222 (2009).
\bibitem{Stander}              N. Stander, B. Huard and D. Goldhaber-Gordon, Phys. Rev. Lett. 102, 026807 (2009).
\bibitem{Chen}                 X. Chen, J.-W. Tao and Y. Ban, Eur. Phys. J. B 79, 203 (2011).
{\bibitem{Blanter} Ya. M. Blanter
and M. B\"uttiker, Physics Reports 336, 1 (2000).}

\end{thebibliography}
\end{document}